\documentclass[aps,preprint,showpacs,amsmath,amssymb,prd,nofootinbib]{revtex4-1}
\usepackage{amssymb,axodraw,graphics,graphicx}

\begin{document}
%\vspace{2.0truecm}
\title
%\begin{center}
{\Large \bf Relating Neutrino Masses to dilepton modes of Doubly Charged Scalars}
%\end{center}

%\vspace{0.9truecm}
\author{Chian-Shu~Chen$^{1,2}$\footnote{e-mail: chianshu@phys.sinica.edu.tw}and
 C.~Q.~Geng$^{3}$\footnote{e-mail: geng@phys.nthu.edu.tw}
 }
\affiliation{$^{1}$Department of Physics, National Cheng Kung University, Tainan, Taiwan 701\\$^{2}$Institute of Physics, Academia Sinica, Taipei, Taiwan 115\\$^{3}$Department of Physics, National Tsing Hua University,
Hsinchu, Taiwan 300}

\date{\today}

%\vspace{2.0truecm}

\begin{abstract}
We study a model with Majorana neutrino masses generated through  doubly charged scalars
at two-loop level.
 We give explicit relationships between  the neutrino masses and
  the same sign dilepton decays of the doubly charged  scalars.
  In particular, we demonstrate  that in the tribimaximal limit of the neutrino mixings, 
  %the lower bound of the lightest neutrino mass as well as
   the absolute neutrino masses and Majorana phases
  can be extracted through the measurements of  
   the dilepton modes at colliders.  

\noindent
\end{abstract}

\maketitle

\section{Introduction}
It has been revealed that at least two light neutrinos have nonzero masses and the mixing matrix is characterized by two large mixing angles from neutrino oscillation experiments. These evidences exhibit new physics beyond the standard model (SM). However, the origin of neutrino masses is still mysterious even though considerable efforts have been put in both theory and experiment for decades. 
In particular, two crucial properties of neutrinos, which can not be disclosed in the neutrino oscillation experiments, are the Majorana nature of neutrinos and the absolute values of the neutrino masses. 
Presently, the only neutrino experiment that could provide a direct evidence of
Majorana neutrinos  is the search for the neutrinoless double beta ($0\nu\beta\beta$) decays,
%. 
which could be used 
to set some limits on the absolute neutrino masses~\cite{Bilenky:2001rz}. 
%The current lower bounds on the half-life measurements of the $0\nu\beta\beta$ decays of the different nuclei are reached in the Heidlberg-Moscow~\cite{Bakalyarov:2003jk}, CUORICINO~\cite{Arnaboldi:2008ds} and other experiments~\cite{pdg}. 
The absolute  neutrino mass can be also  measured through the electron energy spectrum away from the end-point in the nuclear beta decay (i.e. the tritium decay, $m_{\beta} = \sqrt{\sum_{i}|U_{ei}|^2m_{\nu_{i}}^2} < 2$ eV)~\cite{pdg,Otten:2009hg}, while the sum of the neutrino masses 
has been constrained from cosmology, given by
$\sum_{i}m_{\nu_{i}} < 0.58$ eV (95\%~CL)~\cite{WMAP7}.

Instead of probing neutrino properties at low energy experiments, the possibility to explore them
%the mechanism of the neutrino mass generation 
at the Large Hadron Collider (LHC) has been widely studied in the literature.
%many models,
%
%such as the one which employ the 
There are three different types of seesaw mechanisms to generate neutrino masses at tree level.
%Although these models
 %may imply a highly suppressed factor either from a very high mass scale of new particles at
%${\cal O}(10^{14})$ GeV or a very small Yukawa coupling at ${\cal O}(10^{-11})$,
%. 
%it is still possible to 
%discriminate the  mechanisms if 
%
%the new particles are reachable at colliders~\cite{delAguila:2008cj}. 
The ``minimal'' type-$\mathrm{I}$ seesaw mechanism~\cite{Minkowski:1977sc} can not be directly tested at collider experiments due to the suppressed mixings between the light neutrinos and heavy right-handed singlets unless some symmetry is introduced~\cite{Datta:1993nm,Almeida:2000pz,Han:2006ip,del Aguila:2007em}. However, the unsuppressed gauge interactions~\cite{Huitu:1996su,Franceschini:2008pz} of both scalar and fermionic triplets will help us to test the ideas of type-II~\cite{Magg:1980ut} and type-III~\cite{Foot:1988aq} seesaw mechanisms at the LHC, respectively. Furthermore, an one to one correspondence between dilepton decay widths and neutrino masses exists
in the type-II seesaw model where the triplet scalar $T$ couples to the left-handed lepton doublets $\ell_{L}$ via the gauge invariant Yukawa interaction:
\begin{eqnarray}\label{LLT}
L = \bar{\ell^c}_{aL}h_{ab}i\tau_2T\ell_{bL} + h.c. \,,
\end{eqnarray} 
where the indices $a,b$ denote $e, \mu, \tau$ and $\ell_{jL}$  represents the left-handed lepton doublet of the jth flavor.
The neutrino mass matrix is given by 
\begin{eqnarray}\label{type2mass}
m_{\nu ab} = \sqrt{2}h_{ab}v_{T}
\end{eqnarray}
after the triplet receives the vacuum expectation value (VEV), $\langle T \rangle = v_{T}/\sqrt{2}$. The decay width of each same sign dilepton decay mode of the  doubly charged Higgs scalar $P^{\pm\pm}$, $\Gamma(P^{\pm\pm} \rightarrow \ell^{\pm}_{aL}\ell^{\pm}_{bL})$, is directly related to the corresponding  neutrino mass matrix element through Eq.~(\ref{type2mass}). Thus, the discovery of $P^{\pm\pm}$ may 
help us to understand the Majorana nature of  neutrinos,
and by studying the branching ratios of the dilepton channels we may obtain some important informations such as the absolute neutrino masses and Majojrana phases. The related phenomenologies have been extensively studied in the literature~\cite{Garayoa:2007fw}.      

In this work, we study a  model originally proposed in Ref.~\cite{Chen:2006vn},
in which a discrete symmetry is imposed to forbid the Yukawa coupling in Eq.~(\ref{LLT}) at tree level. As a result, the neutrino masses are generated at two-loop level with the normal hierarchy spectrum~\cite{Chen:2007dc},
while the $0\nu\beta\beta$ decay arises at tree level~\cite{SV,Chen:2006vn}. One of the interesting properties of the model is that it shares the same feature of the direct link between the decay widths of the same sign dilepton modes and neutrino masses as the type-II seesaw model. We will focus on the parameters which can not be measured in the neutrino oscillation experiments such as
the absolute masses  of three light neutrinos and 
the Majorana phases $\psi_{1}$ and $\psi_{2}$. By assuming the tribimaximal mixings~\cite{Harrison:2002er} and utilizing the measured 
mass square differences for the neutrinos, we derive some explicit relations between the neutrino masses and
the branching fractions of the doubly charged scalars to the charged lepton pairs. In addition, as
the neutrino masses are proportional to the products of the charged lepton masses $m_{a}m_{b}$ due to the loop integral, the fractions $\frac{BR(P^{\pm\pm}\rightarrow e^{\pm}\mu^{\pm})}{BR(P^{\pm\pm}\rightarrow e^{\pm}\tau^{\pm})} = \frac{m^2_{\tau}}{m^2_{\mu}}$ and $\frac{BR(P^{\pm\pm}\rightarrow \mu^{\pm}\mu^{\pm})}{BR(P^{\pm\pm}\rightarrow \tau^{\pm}\tau^{\pm})} = \frac{m^4_{\tau}}{m^4_{\mu}}$ are
much larger than those of unity  predicted in the type-II seesaw model in the limit of the tribimaximal mixings.
Clearly, we are able to differentiate the two types of the models by counting the events arising from the dilepton decays of the doubly charged Higgs scalars at the LHC. 

Our paper is organized as follows. In Sec. II, we briefly introduce the model. In Sec. III,
we relate  the neutrino masses with  the 
dilepton modes of the doubly charged scalars.  We conclude our results in Sec. IV. 

\section{The model}
The simplest version of the model has been given in Ref.~\cite{Chen:2006vn} with some of its phenomenologies presented in Ref.~\cite{Chen:2007dc}. The idea of the model is to suppress the Yukawa interaction in Eq.~(\ref{LLT}) at tree level and induce it radiatively. Here, we give an explicit example\footnote{Two possible scenarios have been mentioned in the footnote of Ref.~\cite{Chen:2006vn}.} to forbid the tree contributions to the neutrino masses. The model consists two Higgs doublets $\phi_1$ and
$\phi_2$, one  triplet $T$ and one doubly charged singlet $\Psi$ with the hypercharges of $-1/2$, $-1$ and $2$, respectively. 
Besides the gauge symmetry,
there is a $Z_2$ discrete symmetry 
with the transformations of  $\phi_1\rightarrow + \phi_1,\ \phi_2\rightarrow - \phi_2,\
T\rightarrow - T,\ \Psi\rightarrow + \Psi$, and $f\to f$, where $f$ represents the SM fermion.
The most
general potential in this model can be written as
\begin{eqnarray}\label{potential}
V &=& -\mu_1^2\phi_1^{\dag}\phi_1 + \lambda_1(\phi_1^{\dag}
\phi_1)^2 - \mu_2^2\phi_2^{\dag}\phi_2 + \lambda_2(\phi_2^{\dag}
\phi_2)^2 \nonumber \\
&& - \mu_T^2Tr(T^{\dag}T) + \lambda_T[Tr(T^{\dag}T)]^2 +
\lambda'_TTr(T^{\dag}TT^{\dag}T) \nonumber \\
&& + m^2\Psi^{\dag}\Psi + \lambda_{\Psi}(\Psi^{\dag}\Psi)^2
\nonumber \\
&& + \kappa_{\phi_1}Tr(\phi_1^{\dag}\phi_1T^{\dag}T) + \kappa'_
{\phi_1}\phi_1^{\dag}TT^{\dag}\phi_1 + \kappa_{\Psi_1}\phi_1^
{\dag}\phi_1\Psi^{\dag}\Psi \nonumber \\
&& + \kappa_{\phi_2}Tr(\phi_2^{\dag}\phi_2T^{\dag}T) + \kappa'_
{\phi_2}\phi_2^{\dag}TT^{\dag}\phi_2 + \kappa_{\Psi_2}\phi_2^{\dag}
\phi_2\Psi^{\dag}\Psi \nonumber \\
&& + \lambda_{3}\phi_1^{\dag}\phi_1\phi_2^{\dag}\phi_2 + \lambda_{4}\phi_1^{\dag}\phi_2\phi_2^{\dag}\phi_1 + \rho
Tr(T^{\dag}T\Psi^{\dag}\Psi) \nonumber \\
&& + (M\phi_1^TT^{\dag}\phi_2 + \lambda_{5}\phi_1^{\dag}\phi_2\phi_1
^{\dag}\phi_2 + \lambda\tilde{\phi}_1^{\dag}T\tilde{\phi}_2^*\Psi + {\rm H.c.}),
\end{eqnarray}
where
$\tilde{\phi_i} = i\tau_2\phi_i^*$.
Under the symmetries, the doubly charged singlet  $\Psi$ couples
 to the right-handed charged leptons $l_{R}$ via the Yukawa interaction, 
\begin{eqnarray}
\emph{L}_{Y} = Y_{ab}\bar{\ell^{c}}_{aR}\ell_{bR}\Psi + h.c.,  
\end{eqnarray}
where $Y_{ab}$ is a $3\times3$ symmetric matrix with the indices   
%\textbf{$a, b$ stand for flavor indices $e, \mu, \tau$},
$a,b$ stand for $e, \mu, \tau$ and $\ell_{jR}$  represent the right-handed lepton singlets.
We note that the interaction between the scalar triplet and the left-handed lepton doublets is not allowed by imposing the $Z_2$ symmetry such that 
there is no neutrino mass term 
at  tree level unlike  the type-II seesaw mechanism. When the scalar fields $\phi_{1,2}$ and $T$ develop VEVs, both the gauge  and $Z_2$ discrete symmetries  spontaneously break down. 
%Hence, the effective couplings in Eq.~(\ref{LLT}) can be induced, and t
The neutrino masses will be generated through two-loop diagrams as shown in Ref.~\cite{Chen:2006vn}, given by
\begin{eqnarray}\label{numass}
(m_{\nu})_{ab} = \frac{g^4v_TY_{ab}\sin{2\theta}}{\sqrt{2}}m_am_b
\left[ I(M^2_{P_1}) - I(M^2_{P_2})\right],
\end{eqnarray}
where $v_{T} < 4$ GeV is the VEV of the scalar triplet $T$, bounded by the $\rho$-parameter ($=M^2_{W}/M^2_{Z}\cos^2{\theta_{W}}$)~\cite{pdg,Gunion:1990dt}, $m_{a,b}$ stand for the charged lepton masses,
$M_{P_{1,2}}$ are the masses of the doubly charged scalar eigenstates $P_{1,2}^{\pm\pm}$ with the mixing angle $\theta$ defined by 
\begin{eqnarray}\label{mixing}
\left(\begin{array}{c}P_1^{\pm\pm}
\\P_2^{\pm\pm}\end{array}\right) =
\left(\begin{array}{cc}\cos{\theta} & \sin{\theta} \\-\sin{\theta}
& \cos{\theta}\end{array}\right)\left(\begin{array}{c}T^{\pm\pm}
\\\Psi^{\pm\pm}\end{array}\right),
\end{eqnarray} 
and the integral $I(M_{P_i}^2)$ is expressed as 
\begin{eqnarray}
I(M^2_{P_i}) = \int\frac{d^4q}{(2\pi)^4}\int\frac{d^4k}{(2\pi)^4}
\frac{1}{k^2 - m_a^2}\frac{1}{k^2 - M_W^2}\frac{1}{q^2 - M_W^2}
\frac{1}{q^2 - m_b^2}\frac{1}{(k-q)^2 - m_{P_i}^2}\,,
\end{eqnarray}
which can be approximated to be $I \sim \frac{1}{(4\pi)^4M_{P_i}^2}\log^2\left(\frac{M_W^2}{M_{P_i}^2}\right)$  
for $M_{P_i} > M_W$~\cite{TWO}.
%In Appendix A, we show the details of the scalar mixings  of the model.
 From Eq. (\ref{numass}), we see that
 the one to one correspondence between neutrino mass elements $m_{\nu_{ab}}$ and the Yukawa couplings $Y_{ab}$
  provides us the opportunity to determine the neutrino masses through 
  the measurements 
  %studying the phase space 
  of the dilepton modes.
We note that even for small neutrino masses,
the $0\nu\beta\beta$ decays in this model can be large since they are dominated by the exchanges of the doubly charged 
scalars at tree level~\cite{Chen:2006vn}.
  
%The neutrino masses can be generated radiatively.

%

%The neutrino masses can be generated at two-loop level~\cite{Chen:2006vn},
% 
%given by
%\begin{eqnarray}
%(m_{\nu})_{ab} = \frac{g^4v_TY_{ab}\sin{2\delta}}{\sqrt{2}}m_am_b
%\left[ I(M^2_{P_1}) - I(M^2_{P_2})\right],
%\end{eqnarray}
%with the integral $I(M_{P_i}^2)$ defined by
%\begin{eqnarray}
%I(M^2_{P_i}) = \int\frac{d^4q}{(2\pi)^4}\int\frac{d^4k}{(2\pi)^4}
%\frac{1}{k^2 - m_a^2}\frac{1}{k^2 - M_W^2}\frac{1}{q^2 - M_W^2}
%\frac{1}{q^2 - m_b^2}\frac{1}{(k-q)^2 - m_{P_i}^2}.
%\end{eqnarray}
%The integration can be approximated to be $I \sim \frac{1}{(4\pi)^4M_{P_i}^2}\log^2\left(\frac{M_W^2}{M_{P_i}^2}\right)$  
%for $M_{P_i} > M_W$~\cite{TWO}.
%We note that even for small neutrino masses,
%the $0\nu\beta\beta$ decays in this model can be large as they are dominated through the exchanges of the doubly charged scalars at tree level~\cite{Chen:2006vn}.
%, 

\section{Decay Branching ratios related to  neutrino masses}

\subsection{Dilepton decays}

There are several channels that the doubly charged scalars
can decay into, such as $P^{\pm\pm} \rightarrow \ell_{aR}^{\pm}\ell^{\pm}_{bR}, W^{\pm}W^{\pm}, W^{\pm}P^{\pm}, P^{\pm}P^{\pm}$ and $W^{\pm}W^{\pm}P^{0}$,
where $P^{\pm\pm}$ are referred to the lighter mass eigenstate among
$P^{\pm\pm}_i\ (i=1,2)$. Due to the kinematical consideration, we expect that the first two kinds of the modes contribute the most part to the width of $P^{\pm\pm}$ if $M_{P^{\pm\pm}}\sim M_{P^{\pm}}\sim M_{P^0}$. However, as long as the mass splitting between $P^{\pm\pm}$ and $P^{\pm}$($M_{P^0}$) is large enough, the last three channels may dominate at the high mass region.
In our discussion, we assume that the last three types of decay channels with scalar(s) in the final states are not allowed. Hence, the two-loop suppression factor in neutrino masses of Eq.~(\ref{numass}) makes the fraction of $\frac{\Gamma(l_{aR}l_{bR})}{\Gamma(WW)} \approx \frac{16(4\pi)^8}{g^{12}}\left(\frac{|m_{\nu_{ab}}|M_{P_1}}{m_am_b}\right)^2\left(\frac{M_{W}}{v_{T}}\right)^4\sim{\cal O}(10^{20})$, so that the branching ratio with the dilepton final states is almost $100\%$ for
% with all regime of the parameter space of
 $v_{T} \lesssim {\cal O}(1)$ GeV by taking $m_{\nu_{ab}} = 0.1$ eV and $M_{P_1} = 200$ GeV. %In particular, the $e^{\pm}e^{\pm}$ modes are dominant in the final sates. 
%On the other hand, 
While the fraction in the seesaw type-II model is $(\frac{m_{\nu_{ab}}}{M_{H^{\pm\pm}}})^2(\frac{M_{W}}{v_{T}})^4 \gtrsim 1$ only for $v_{T} \lesssim {\cal O}(10^{-4})$ GeV with the same input.
%However, in the type-II seesaw model,  
%the decay branching ratios of $P^{\pm\pm}\to \ell^{\pm}\ell^{\pm}$
 %are larger than those of $W^{\pm}W^{\pm}$ only for $v_{T} < {\cal O}(10^{-4})$ GeV . 
 We take the values of $m_{\nu_{ee}} \sim 0.01$ eV,
   $m_{\nu_{e\mu}} \sim m_{\nu_{e\tau}} \sim 0.1$ eV,
   and $m_{\nu_{\mu\mu}} \sim m_{\nu_{\mu\tau}} \sim m_{\nu_{\tau\tau}} \sim 1$ eV to illustrate the possible branching ratios of the dilepton modes in our model.
  These values are based on the texture of the neutrino mass matrix with the normal hierarchical spectrum as predicted by this model~\cite{Chen:2007dc}.
% \begin{eqnarray}
% m_{\nu_{NH}} = m\left(\begin{array}{ccc}\delta & \epsilon & \epsilon \\\epsilon & 1 + \eta & 1 + \eta \\\epsilon & 1 + \eta & 1 + \eta\end{array}\right)
% \end{eqnarray}
  The estimations of the decays
  for $M_{P} = 200$ GeV and $v_{T} = 1$ GeV are shown in Table~\ref{tab:BR}.
  One should keep in mind that there might exist some cancellations among the combinations of three neutrino masses and Majorana phases in each element of the neutrino mass matrix. For example, the component $m_{\nu_{ee}}$ may go to zero with certain values of Majoran phases and the lightest neutrino mass for the normal hierarchical spectrum. In this case, $0\nu\beta\beta$ decays may be out of the reach of the current experimental sensitivity. Therefore, the branching ratios of the dilepton modes shown in Table~\ref{tab:BR} will change drastically with different values of Majorana phases and neutrino masses,
  which will be discussed in Sec. III C.
  %  .

The production of $P^{\pm\pm}$ is dominated by the Drell-Yan process, for which the next to leading-order contribution from QCD enhances the cross section by a factor of 1.25 at the LHC~\cite{Muhlleitner:2003me}. Several simulations have been performed for $P^{\pm\pm}\rightarrow \ell^{\pm}_{a}\ell^{\pm}_{b}$ at the LHC~\cite{Azuelos:2005uc}. For channels of the doubly charged Higgs decays into $e$ or $\mu$, the SM background is shown to be negligible in the signal region of the high invariant mass close to $M_{P^{\pm\pm}}$. In contrast, the spectrum of the missing transverse momentum for the $\tau$ decay product will be softer due to the subsequent decays of $\tau \rightarrow e\nu\bar{\nu}$ and $\tau \rightarrow \mu\nu\bar{\nu}$ with a branching ratio around 17\% in each channel. Additional backgrounds with jets, such as $W^{\pm}W^{\pm}jj$,   may fake the hadronic $\tau$ decays. As a result, the $\tau$ tag efficiency is about 50\% and the fake rate is around 1\%~\cite{Atlas}. The detector-specific error analysis is beyond the scope of this paper. We have used uniform uncertainties for all branching ratios for the rough estimation of the effect. For example, the approximate event numbers for the pair production of $P^{\pm\pm}$ are 15000, 3000 and 900 with $M_{P} = 200, 300$ and $400$ GeV, respectively, at the LHC ($\sqrt{s} = 14$ TeV) for the integrated luminosity of ${\it L} = 300 fb^{-1}$, $BR(P^{\pm\pm}\rightarrow \ell_a^{\pm}\ell_b^{\pm}) = 100\%$, and the  efficiency $\epsilon_{eff} = 0.5$. We write the dilepton decay widths as  
\begin{eqnarray}
\Gamma(P^{\pm\pm} \rightarrow \ell_{aR}^{\pm}\ell_{bR}^{\pm})&=&
\frac{|Y_{ab}|^2}{8\pi(1+\delta_{ab})}s^2_{\theta}M_{P}.
\label{eq:13}
\end{eqnarray}
\begin{table}[th]
\caption{\label{tab:BR}The branching ratios of  $P^{\pm\pm}\to \ell_a^{\pm}\ell_b^{\pm}$  ($\ell_{a,b}^\pm= e^\pm, \mu^\pm$ and $\tau^\pm$) by assuming the neutrino mass elements of $m_{\nu_{ee}}$ = 0.01, $m_{\nu_{e\mu}} = m_{\nu_{e\tau}}$ = 0.1, and $m_{\nu_{\mu\mu}} = m_{\nu_{\mu\tau}} = m_{\nu_{\tau\tau}}$ = 1 eV for $v_{T}$ = 1 GeV and $M_{P}$ = 200 GeV~\cite{Chen:2007dc}.}\begin{center}
\begin{tabular}{|c|c|c|c|c|c|} \hline \hline
$BR_{e^{\pm}e^{\pm}}$ & $BR_{e^{\pm}\mu^{\pm}}$ & $BR_{e^{\pm}\tau^{\pm}}$ & $BR_{\mu^{\pm}\mu^{\pm}}$ & $BR_{\mu^{\pm}\tau^{\pm}}$ & $BR_{\tau^{\pm}\tau^{\pm}}$ \\ \hline 
$0.995$  & $4.6\times10^{-3}$ & $1.6\times10^{-5}$ & $5.4\times10^{-6}$ & $3.8\times10^{-8}$ & $6.7\times10^{-11}$ \\ 
\hline \hline
\end{tabular}
\end{center}
\end{table} 

\subsection{
Neutrino masses and mixings}
%Combining with the neutrino masses in Eq.~(\ref{numass}) it is interesting to
%note that 
Since the decay widths in Eq.~(\ref{eq:13})
are proportional to $Y_{ab}$, which are related to the neutrino masses in Eq.~(\ref{numass}) it provides with us an opportunity to study the spectrum of neutrinos at the LHC. We can express the branching ratios of the same sign charged lepton pair modes in terms of the components of the neutrino mass matrix  
\begin{eqnarray}\label{BRs}
BR_{ab} = \frac{\Gamma(\ell^{\pm}_{aR}\ell^{\pm}_{bR})}{\Gamma_{total}} &=& \frac{\sin^2_{\theta}M_{P1}}{\Gamma_{total}\times4\pi g^8v^2_{T}\sin^2{2\theta}\left[I(M^2_{P_1} - I(M^2_{P_2}))\right]}\times\frac{|m_{\nu_{ab}}|^2}{(1 + \delta_{ab})m^2_{a}m^2_{b}} \nonumber \\
&\propto& \frac{|m_{\nu_{ab}}|^2}{(1 + \delta_{ab})m^2_{a}m^2_{b}}.
\end{eqnarray}
%r. 
It will be clear later that the overall factor including the total decay width is irrelevant and the dependence of charged lepton masses appearing in the dilepton branching ratios is due to the loop integral in our model. 
 
Similar to the CKM mixing matrix in the quark sector, the neutrino mass matrix can be diagonalized by the unitary matrix $U_{PMNS}$,
defined by~\cite{Maki:1962mu}, 
\begin{eqnarray}
M_{\nu} = U_{PMNS}\left(\begin{array}{ccc}m_1 & 0 & 0 \\0 & m_2e^{i\psi_1} & 0 \\0 & 0 & m_3e^{i\psi_2}\end{array}\right)U^{T}_{PMNS},
\end{eqnarray} 
where $\psi_{1,2}$ are referred to as the Majorana phases and the PMNS matrix can be parametrized as 
\begin{eqnarray}
U_{PMNS} = \left(\begin{array}{ccc}c_{12}c_{13} & s_{12}c_{13} & s_{13}e^{-i\delta} \\-s_{12}c_{23} - c_{12}s_{23}s_{13}e^{i\delta} & c_{12}c_{23} - s_{12}s_{23}s_{13}e^{i\delta} & s_{23}c_{13} \\s_{12}s_{23} - c_{12}c_{23}s_{13}e^{i\delta} & -c_{12}s_{23} - s_{12}c_{23}s_{13}e^{i\delta} & c_{23}c_{13}\end{array}\right)
\end{eqnarray}
with  the Dirac phase of $\delta$. Currently, the neutrino oscillation experiments can only give the mass-square differences~\cite{pdg}:
\begin{eqnarray}
\Delta m^2_{21} = (7.59\pm0.20) \times10^{-5} {\rm eV^2}, \quad |\Delta m^2_{32}| = (2.43\pm0.13) \times10^{-3} {\rm eV^2},  
\end{eqnarray}
and the mixing angles ~\cite{pdg}:
\begin{eqnarray}
\sin^2{(2\theta_{12})} = 0.87\pm0.03, \quad \sin^2{(2\theta_{23})} \simeq 1, \quad \sin^2{(2\theta_{13})} < 0.19.   
\end{eqnarray}
%. 
A very good approximation of leptonic mixing matrix is proposed with the so-called tribimaximal mixing form~\cite{Harrison:2002er} 
\begin{eqnarray}
\sin^2{\theta_{12}} = \frac{1}{3}, \quad \sin^2{\theta_{23}} = \frac{1}{2}, \quad \sin^2{\theta_{13}} = 0. 
\end{eqnarray}
In this limit, we can express the elements of the neutrino mass matrix as 
\begin{eqnarray}\label{masselement}
|m_{\nu_{ee}}|^2 &=& \frac{4}{9}m_1^2 + \frac{4}{9}m_1m_2\cos{\psi_1} + \frac{1}{9}m_2^2\,,\nonumber \\
|m_{\nu_{e\mu}}|^2 &=& \frac{1}{9}m_1^2 - \frac{2}{9}m_1m_2\cos{\psi_1} + \frac{1}{9}m_2^2 \,,\nonumber \\
|m_{\nu_{e\tau}}|^2 &=& |m_{\nu_{e\mu}}|^2 \,,
\nonumber \\
|m_{\nu_{\mu\mu}}|^2 &=& \frac{1}{36}m_1^2 + \frac{1}{9}m_2^2 + \frac{1}{4}m_3^2 + \frac{1}{9}m_1m_2\cos{\psi_1} + \frac{1}{6}m_1m_3\cos{\psi_2} + \frac{1}{3}m_2m_3\cos{(\psi_1 - \psi_2)} 
\,, 
\nonumber \\
|m_{\nu_{\mu\tau}}|^2 &=& \frac{1}{36}m_1^2 + \frac{1}{9}m_2^2 + \frac{1}{4}m_3^2 + \frac{1}{9}m_1m_2\cos{\psi_1} - \frac{1}{6}m_1m_3\cos{\psi_2} - \frac{1}{3}m_2m_3\cos{(\psi_1 - \psi_2)}\,, \nonumber \\
|m_{\nu_{\tau\tau}}|^2 &=& |m_{\nu_{\mu\mu}}|^2.
\label{eq:20}
\end{eqnarray}

\subsection{Relations}
Since the normal hierarchical mass spectrum of the light neutrinos is predicted in our model~\cite{Chen:2007dc}, we can parametrize
 the eigenvalues of the masses in terms of the mass differences and the lightest one $m_1$,
 given by
 \begin{eqnarray}
m_1, \quad m_2 = \sqrt{m_1^2 + \Delta m^2_{21}}, \quad m_3 = \sqrt{m_1^2 + \Delta m^2_{21} + \Delta m^2_{32}}. 
\end{eqnarray}
By utilizing Eqs.~(\ref{BRs}) and (\ref{masselement}), we can define the quantity 
\begin{eqnarray}\label{C1}
C_1 = \frac{m^2_{\mu}(2m^2_{\mu}BR_{\mu\mu} + m^2_{\tau}BR_{\mu\tau} + m^2_{e}BR_{e\mu})}{2m^2_{e}(m^2_{e}BR_{ee} + m^2_{\mu}BR_{e\mu})} = \frac{m_1^2 + \frac{5}{6}\Delta m^2_{21} + \frac{1}{2}\Delta m^2_{32}}{m_1^2 + \frac{1}{3}\Delta m^2_{21}}
\end{eqnarray}
such that we can determine the lightest neutrino mass $m_1$ via the quantity $C_{1}$ by measuring the branching ratios of the  $ee, e\mu, \mu\mu$ and $\mu\tau$ channels with the relation 
\begin{eqnarray}
m_1^2 = \frac{(\frac{5}{6} - \frac{1}{3}C_1)\Delta m^2_{21} + \frac{1}{2}\Delta m^2_{32}}{C_1 -1}. 
\end{eqnarray}
Similarly, we can express the masses $m_{2}$ and $m_{3}$ as
\begin{eqnarray}
m^2_2 = \frac{(C_2 + 2)\Delta m^2_{21} + 3C_2\Delta m^2_{32}}{3(1 - C_2)}
\end{eqnarray}
and 
\begin{eqnarray}
m^2_3 = \frac{(9 - 8C_3)\Delta m^2_{32} + (6 - C_3)\Delta m^2_{21}}{9 - 17C_3},
\end{eqnarray}
with the quantities $C_{2}$ and $C_{3}$, defined by
\begin{eqnarray}
C_2 = \frac{m^2_{e}(m^2_{e}BR_{ee} + m^2_{\mu}BR_{e\mu})}{2m^4_{\mu}BR_{\mu\mu} + m^2_{\mu}m^2_{\tau}BR_{\mu\tau} - m^4_{e}BR_{ee}}
\end{eqnarray}
and
\begin{eqnarray}\label{C3}
C_3 = \frac{m^2_{\mu}(2m^2_{\mu}BR_{\mu\mu} + m^2_{\tau}BR_{\mu\tau}) + m^2_{e}(2m^2_{\mu}BR_{e\mu} - m^2_{e}BR_{ee})}{m^2_{e}(m^2_{e}BR_{ee} + m^2_{\mu}BR_{e\mu})}
% = \frac{m^2_3 - \Delta m^2_{32} - \frac{2}{3}\Delta m^2_{21}}{\frac{17}{9}m^2_3 - \frac{8}{9}\Delta m^2_{32} - \frac{1}{9}\Delta m^2_{21}}, \nonumber \\
\end{eqnarray}
respectively. As a result,
 we are able to indirectly determine the neutrino masses 
 by measuring the quantities $C_1$, $C_2$, and $C_3$ in the tribimaximal limit of the neutrino mixings. One can see that the different terms in numerators of $C_{i} ($i =1-3$)$ are all of the same order of magnitude since the large differences in the branching ratios (see Table~\ref{tab:BR}) are compensated by the large differences of the masses of the charged leptons. The degree of accuracy in measuring $C_{i}$ depends on the assumption of a sufficient number of like-sign leptons to be observed, so the sensitivity depends on the number of $P^{\pm\pm}$ and the branching ratios of dilepton modes. 
%We note that a single like-sign dilepton production with a negligible SM background at the LHC has been studied in Refs.~\cite{Huitu:1996su, Allanach:2006fy}, in which it has been shown that the 2-lepton events are about $3$ times larger than the $4$-lepton ones for $200$ GeV $< m_{P^{\pm\pm}} < 400$ GeV. Therefore, it is more effective to probe the dilepton modes with small branching ratios through the search of $P^{\pm\pm} \rightarrow l_{a}^{\pm}l_{b}^{\pm}$. 
 
\begin{figure}[t]
  \centering
    \includegraphics[width=0.49\textwidth]{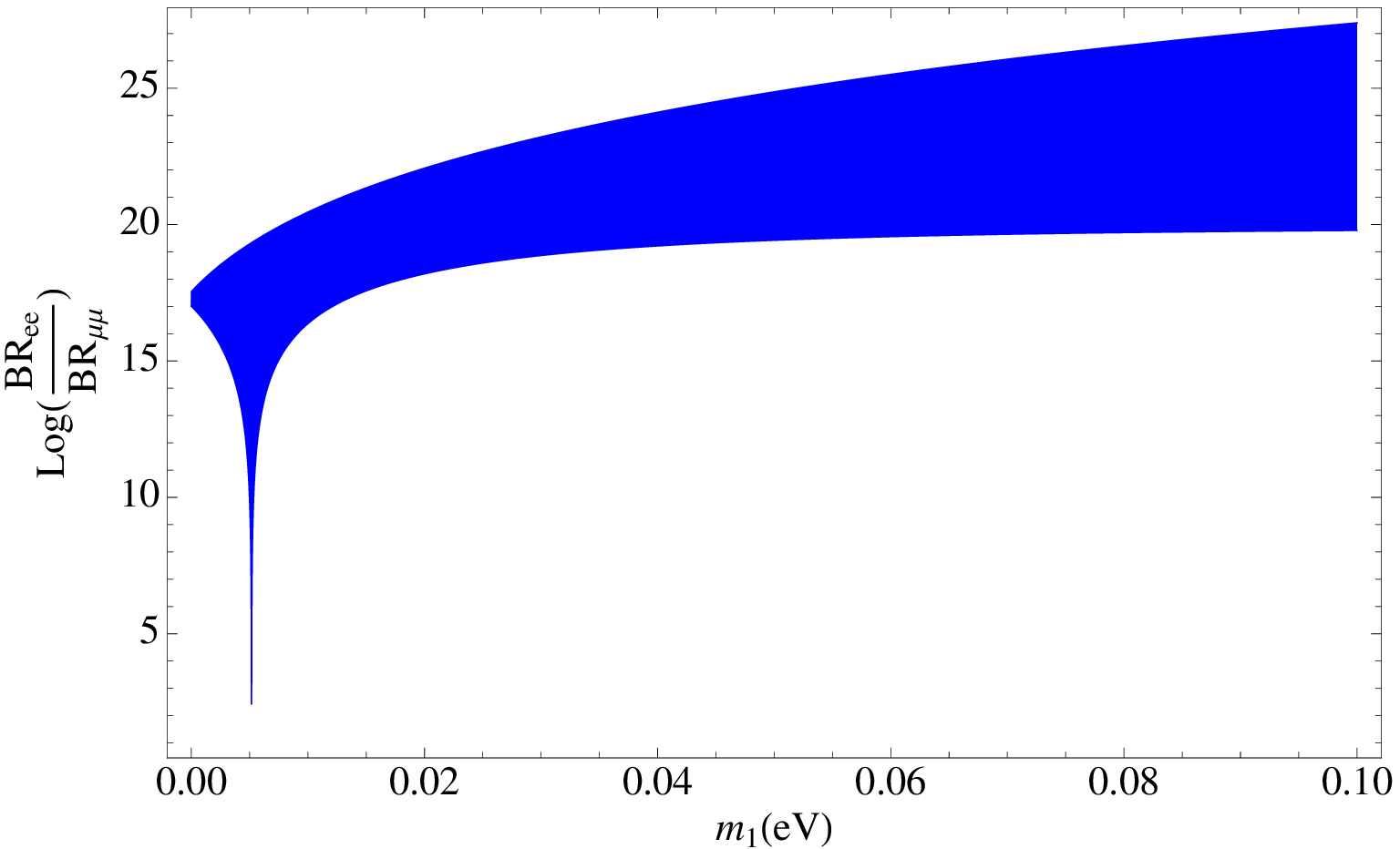}
    \includegraphics[width=0.49\textwidth]{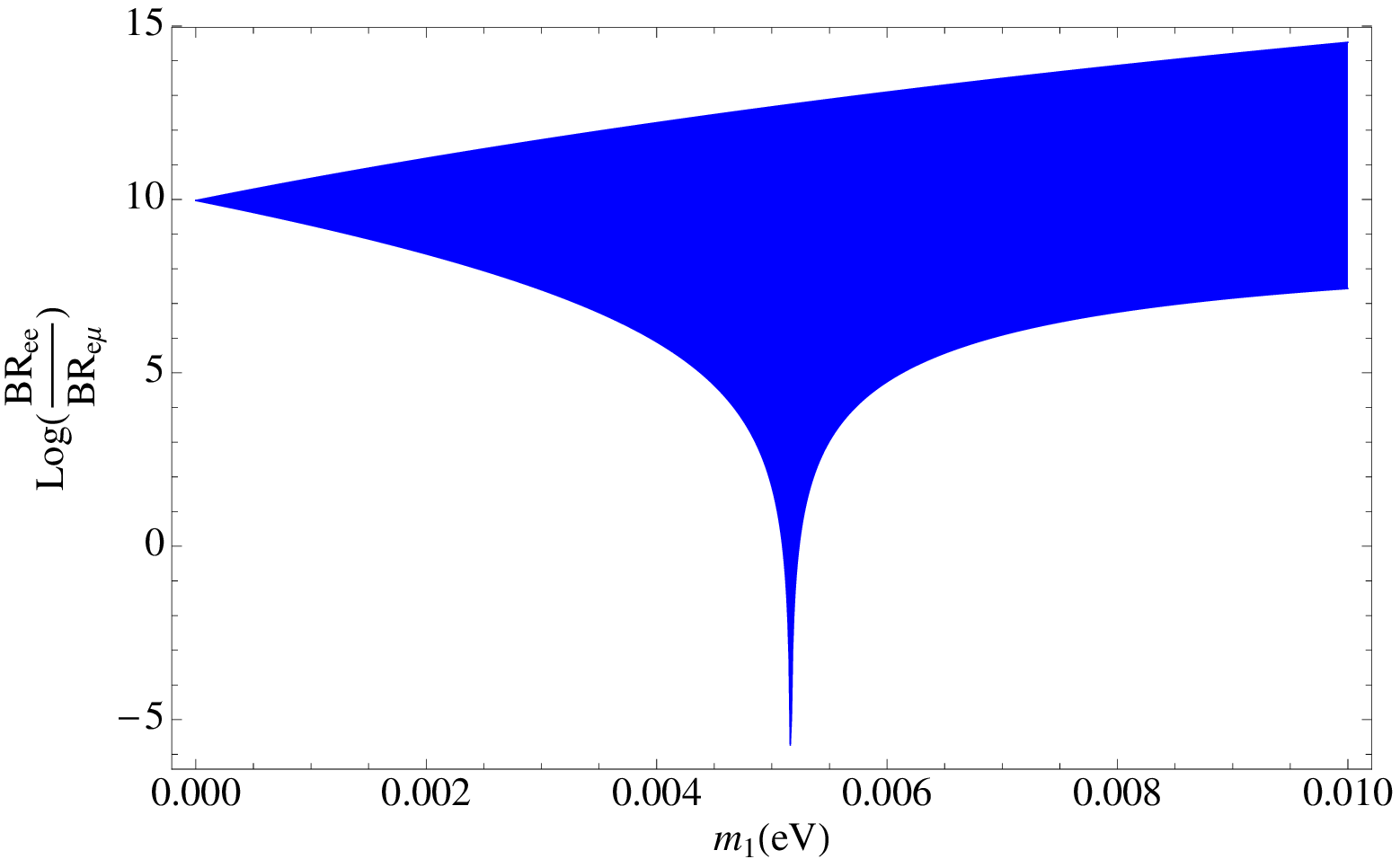}
  \caption{Ratios of $BR_{ee}/BR_{\mu\mu}$ (left) and $BR_{ee}/BR_{e\mu}$ (right) versus 
  the lightest neutrino mass $m_1$ with scanning over the possible values of Majorana phases, where the 
  shadow areas are the allowed regions.
  }
  \label{fig:BRee-uu}
\end{figure}

\begin{figure}[t]
  \centering
    \includegraphics[width=0.49\textwidth]{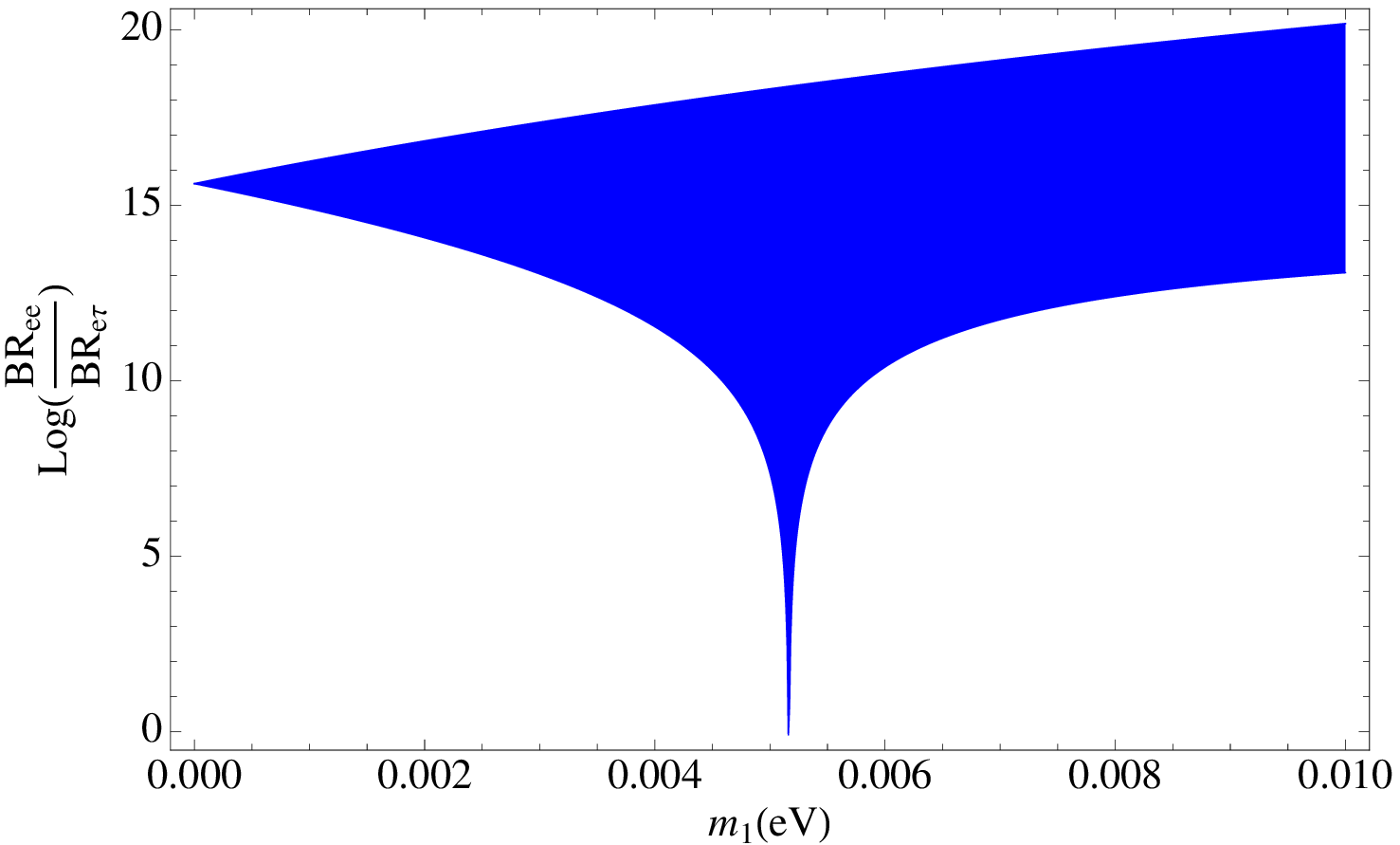}
    \includegraphics[width=0.49\textwidth]{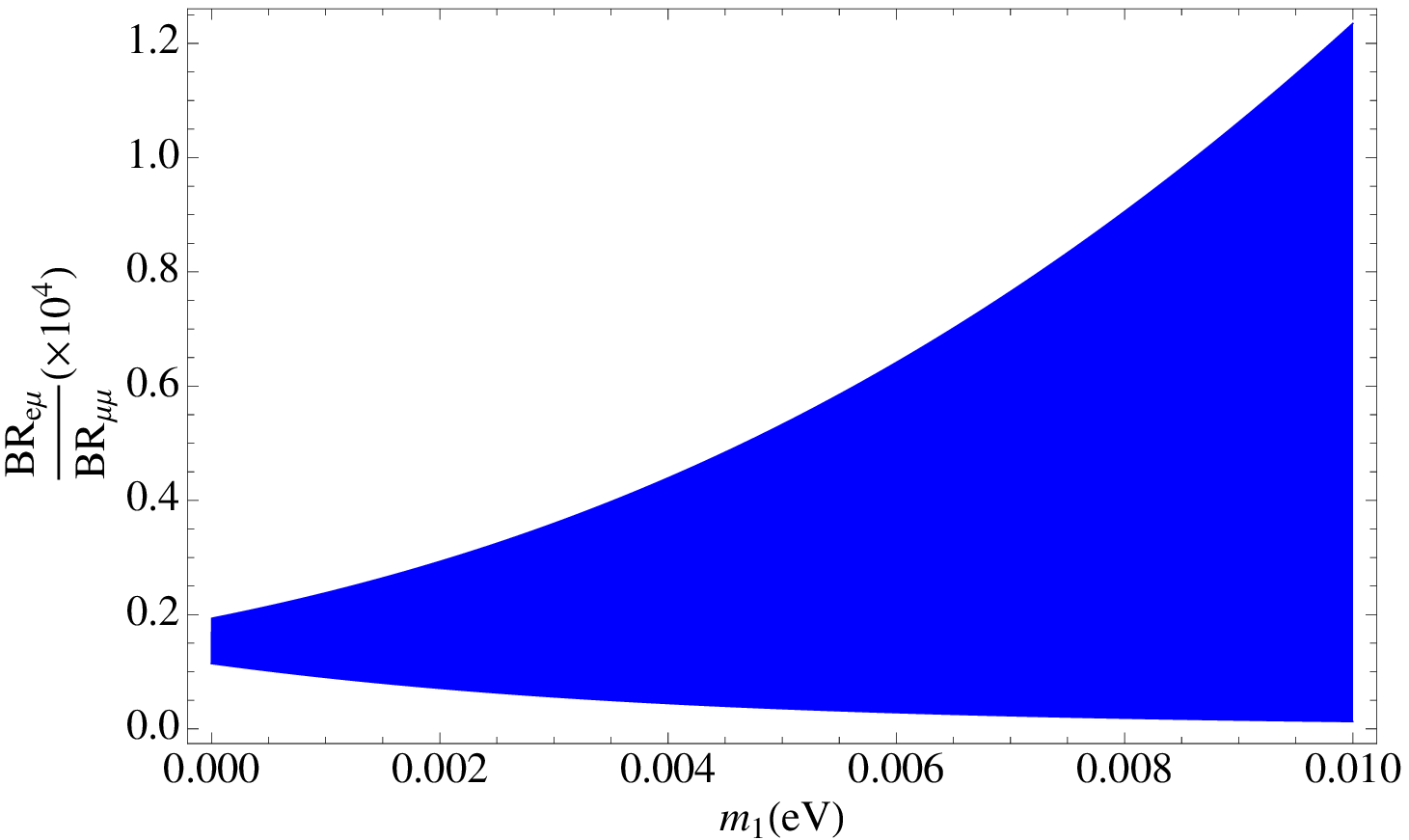}
  \caption{Legend is the same as Fig.~\ref{fig:BRee-uu} but with
  the ratios of $BR_{ee}/BR_{e\tau}$ (left) and $BR_{e\mu}/BR_{\mu\mu}$ (right).}
%   .}
  \label{fig:BRee-et}
\end{figure}

\begin{figure}[t]
  \centering
    \includegraphics[width=0.47\textwidth]{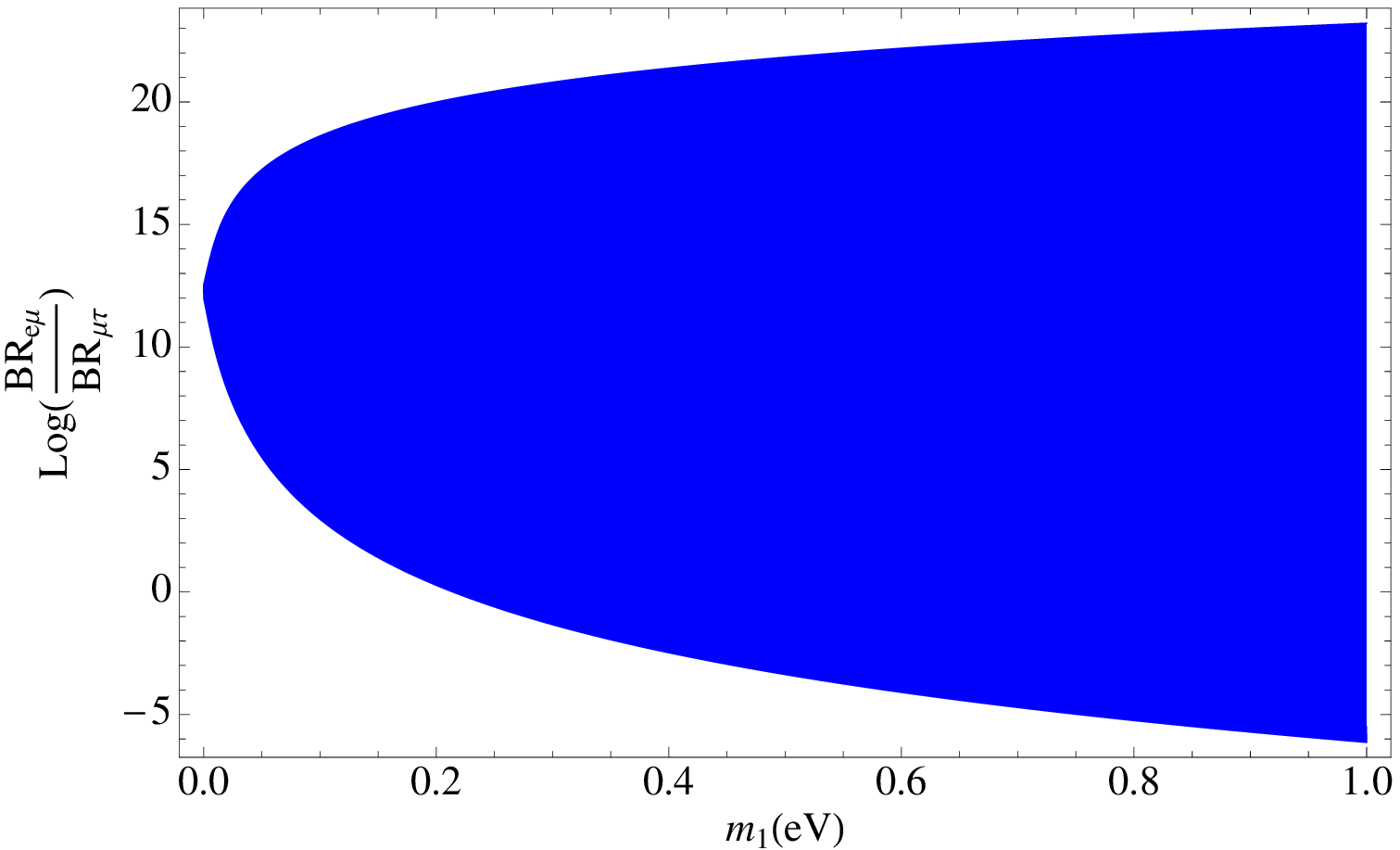}
    \includegraphics[width=0.48\textwidth]{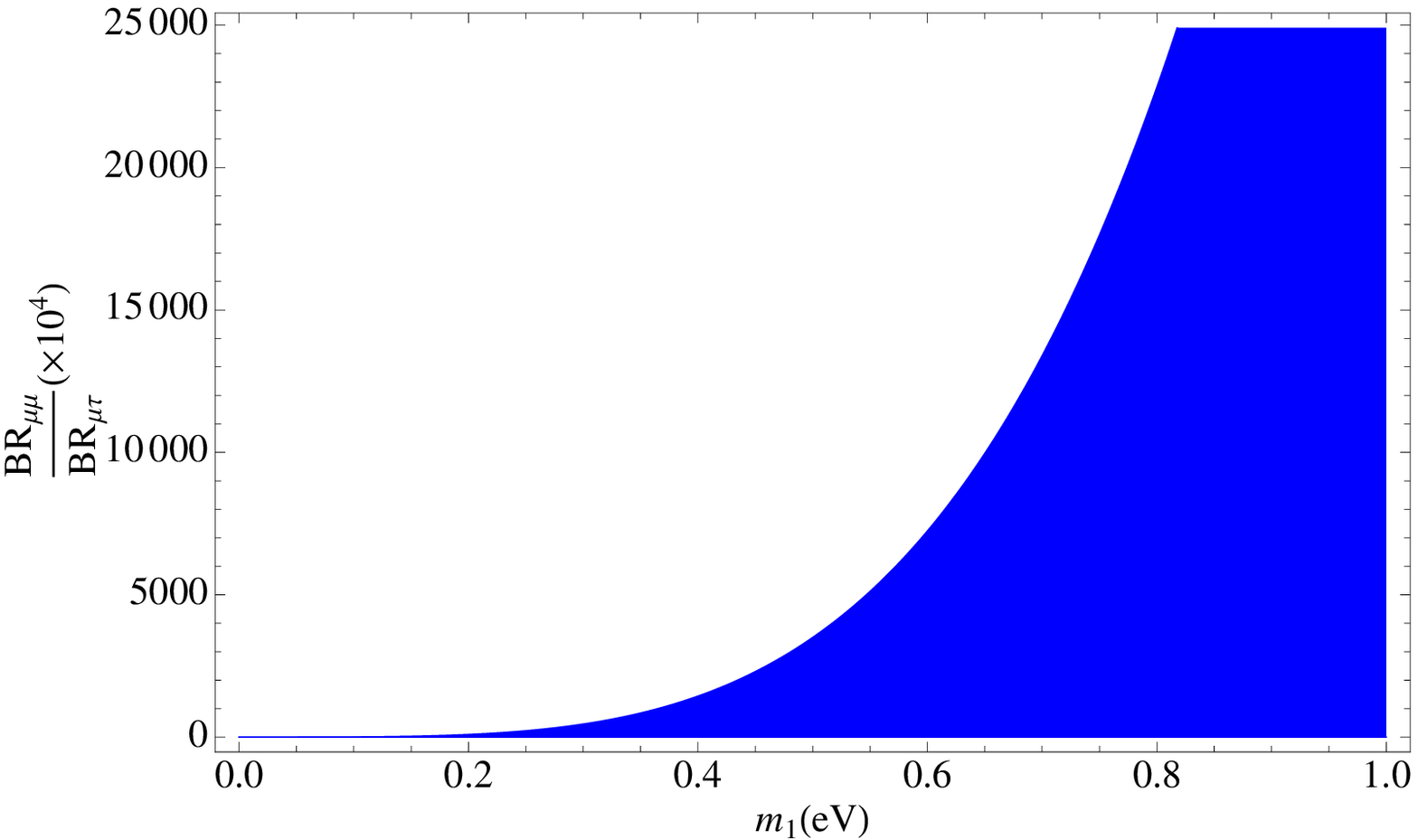}
  \caption{
  Legend is the same as Fig.~\ref{fig:BRee-uu} but with
  the ratios
  of $BR_{e\mu}/BR_{\mu\tau}$ (left) and $BR_{\mu\mu}/BR_{\mu\tau}$ (right).}
  % }
  \label{fig:BReu-uu}
\end{figure}

\begin{figure}[h]
  \centering
    \includegraphics[width=0.48\textwidth]{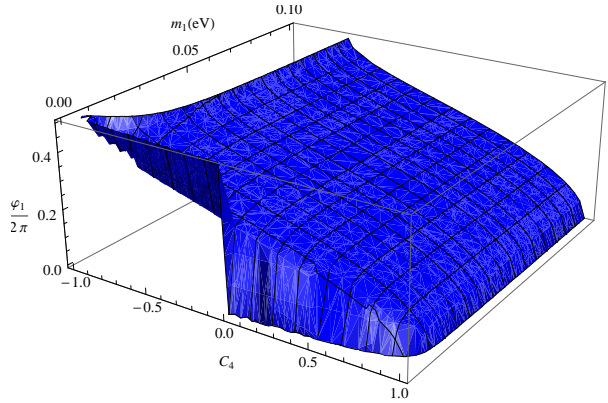}
  \caption{The Majorana phase $\psi_1$  in terms of $C_{4} = \frac{2m^2_{e}BR_{ee} - m^2_{\mu}BR_{e\mu}}{2m^2_{e}BR_{ee} + m^2_{\mu}BR_{e\mu}}$ and $m_1$.}
  \label{fig:psi_1}
\end{figure}

\begin{figure}[h]
  \centering
    \includegraphics[width=0.45\textwidth]{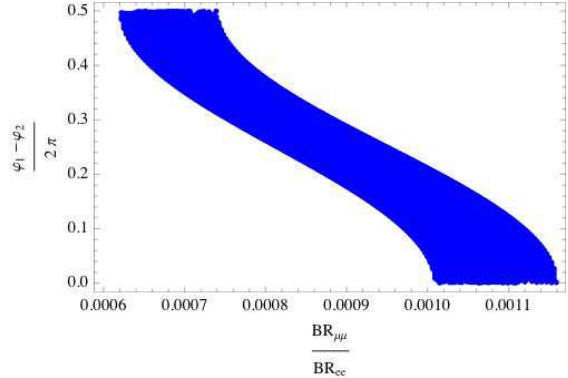}
  \caption{The relative Majorana phase $\psi_1-\psi_{2}$ versus $\frac{BR_{e\mu}}{BR_{\mu\mu}}$ for $m_1\to 0$.}
  \label{fig:psi_2}
\end{figure}

As we mentioned in Sec.~III A, the branching ratios of the dilepton channels are sensitive to the values of
$\psi_{1,2}$ and $m_1$ as given in Eq.~(\ref{eq:20}). Thus, $BR_{l^{\pm}_{a}l^{\pm}_{b}}$ will be very different from those shown in Table~\ref{tab:BR} with different values of $\psi_{1,2}$ and $m_{1}$. However, we would use the ratio 
\begin{eqnarray}\label{BRee}
\frac{BR_{ee}}{BR_{e\mu}} = \frac{1}{2}\frac{m^2_{\mu}}{m^2_{e}}\frac{|m_{\nu_{ee}}|^2}{|m_{\nu_{e\mu}}|^2} = \frac{1}{2}\frac{m^2_{\mu}}{m^2_{e}}\frac{\frac{5}{9}m_1^2 + \frac{4}{9}m_1\sqrt{m_1^2 + \Delta m^2_{21}}\cos{\psi_1} + \frac{1}{9}\Delta m^2_{21}}{\frac{2}{9}m_1^2 - \frac{2}{9}m_1\sqrt{m_1^2 + \Delta m^2_{21}}\cos{\psi_1} + \frac{1}{9}\Delta m^2_{21}}
\end{eqnarray}
to pin down the allowed parameter region as a function of the lightest neutrino mass $m_{1}$. This is demonstrated  in Fig.~\ref{fig:BRee-uu} (right). Similar results of other ratios, such as $\frac{BR_{ee}}{BR_{\mu\mu}}$, ($\frac{BR_{ee}}{BR_{e\tau}}$, $\frac{BR_{e\mu}}{BR_{\mu\mu}}$), and ($\frac{BR_{e\mu}}{BR_{\mu\tau}}$,  $\frac{BR_{\mu\mu}}{BR_{\mu\tau}}$)
%\begin{eqnarray}
%\frac{BR_{ee}}{BR_{\mu\mu}} = \frac{m^2_{\mu}}{m^2_{e}}\frac{\frac{5}{9}m^2_{1} + \frac{4}{9}m_{1}\sqrt{m^2_1 + \Delta m^2_{21}}\cos{\psi_1} + \frac{1}{9}\Delta m^2_{21}}{\frac{7}{18}m^2_1 + \frac{13}{36}\Delta m^2_{21} + \frac{1}{4}\Delta m^2_{32} + m_1(\frac{1}{9}\sqrt{m^2_1 + \Delta m^2_{21}}\cos{\psi_1} + \frac{1}{6}\sqrt{m^2_1 + \Delta m^2_{32} + \Delta m^2_{21}}\cos{\psi_2}) + }
%\end{eqnarray}
are displayed  in Figs.~\ref{fig:BRee-uu} (left),~\ref{fig:BRee-et} (left, right), and \ref{fig:BReu-uu}
(left, right), respectively. It is interesting to note that the branching ratio of $P^{\pm\pm} \rightarrow e^{\pm}e^{\pm}$ reduces significantly in the small region around $m_1\sim0.005$ eV due to the cancellations. In this region, the branching ratios of the rest dilepton modes will be enhanced (see Eq.~(\ref{BRee}), Figs.~\ref{fig:BRee-uu}, and Fig.~\ref{fig:BRee-et}). %Then we will probe the absolute scale $m_{i} (i = 1-3)$ in different ways with different parameter regions. For the case of the cancellation happens in the $m_{\nu_{ee}}$ component, that is $m_1\sim0.005$ eV with certain value of $\psi_{1}$ (see Eq.~(\ref{BRee}), Figs.~\ref{fig:BRee-uu}, and Fig.~\ref{fig:BRee-et}), $BR_{ee}$ will be suppressed and hence branching ratios of the rest dilepton modes will enhance. 
Our relations of Eqs.~(\ref{C1})-(\ref{C3}) will be practically implemented for the lowest mass of the doubly charged Higgs scalars. If there is no cancellation in $m_{\nu_{ee}}$, we can still try to narrow down the parameter space of $m_{1}$ and $\psi_{1,2}$ by measuring the fractions of the dilepton modes. For instance, in the limit of $m_1 \rightarrow 0$ the ratio of $BR_{ee}/BR_{e\mu}$ becomes 
\begin{eqnarray}
\frac{BR_{ee}}{BR_{e\mu}} = \frac{m^2_{\mu}}{2m^2_{e}}.
\end{eqnarray}  
On the other hand, if $m_1$ is measured, the  Majorana phase $\psi_1$ can be extracted from the relation 
\begin{eqnarray}\label{psi1}
\cos{\psi_{1}} = \frac{(7C_4 - 3)m^2_1 + 2C_4\Delta m^2_{21}}{(6 - 2C_4)m_1\sqrt{m^2_1 + \Delta m^2_{21}}},
\end{eqnarray}
where $C_{4}$ is expressed as 
\begin{eqnarray}
C_4 = \frac{2m^2_eBR_{ee} - m^2_{\mu}BR_{e\mu}}{2m^2_eBR_{ee} + m^2_{\mu}BR_{e\mu}}.
\end{eqnarray}
In Fig.~\ref{fig:psi_1}, we plot the allowed region of $\psi_1$ versus 
 $\frac{BR_{ee} - BR_{e\mu}}{BR_{ee} + BR_{e\mu}}$ and  $m_1$. Note that the phase $\psi_1$ becomes indefinite in Eq.~(\ref{psi1}) if we set $m_1 \rightarrow 0$. This is because when the lightest neutrino mass is zero, there is only one Majorana phase left, related to the relative phase of $\psi_1-\psi_2$ as shown in Eq. (\ref{eq:20}).
In this limit, we  obtain  
\begin{eqnarray}\label{Mphase}
\cos{(\psi_1 - \psi_2)} = \frac{\frac{2}{3}(\frac{BR_{\mu\mu}}{BR_{e\mu}})\frac{m^2_{\mu}}{m^2_e}\Delta m^2_{21} - (\frac{13}{12}\Delta m^2_{21} + \frac{3}{4}\Delta m^2_{32})}{\Delta m_{21}\sqrt{\Delta m^2_{21} + \Delta m^2_{32}}}.
\end{eqnarray}
The allowed region  of $\psi_1 -\psi_{2}$ for $m_1\to 0$ 
is displayed in Fig.~\ref{fig:psi_2}, where we have taken
 the uncertainties  of the mass differences from the 
 the solar and atmospherical data \cite{pdg}. 

%After combining other future experiments from oscillations, tritium decays, $0\nu\beta\beta$, and cosmology survey will extract
%
%$m_1$ 
%
%from  other ratios of the dilepton channels.

%to obtain the lower bound on $m_1$ as shown in Fig.~\ref{fig:BRee-uu} (right) where the shadow area is the allowed region by \textbf{scanning all possible values of} the Majorana phase $\psi_{1}$.

%Similar results of other ratios, such as $\frac{BR_{ee}}{BR_{\mu\mu}}$, ($\frac{BR_{ee}}{BR_{e\tau}}$, $\frac{BR_{e\mu}}{BR_{\mu\mu}}$), and ($\frac{BR_{e\mu}}{BR_{\mu\tau}}$,  $\frac{BR_{\mu\mu}}{BR_{\mu\tau}}$)
%\begin{eqnarray}
%\frac{BR_{ee}}{BR_{\mu\mu}} = \frac{m^2_{\mu}}{m^2_{e}}\frac{\frac{5}{9}m^2_{1} + \frac{4}{9}m_{1}\sqrt{m^2_1 + \Delta m^2_{21}}\cos{\psi_1} + \frac{1}{9}\Delta m^2_{21}}{\frac{7}{18}m^2_1 + \frac{13}{36}\Delta m^2_{21} + \frac{1}{4}\Delta m^2_{32} + m_1(\frac{1}{9}\sqrt{m^2_1 + \Delta m^2_{21}}\cos{\psi_1} + \frac{1}{6}\sqrt{m^2_1 + \Delta m^2_{32} + \Delta m^2_{21}}\cos{\psi_2}) + }
%\end{eqnarray}
%are displayed  in Figs.~\ref{fig:BRee-uu} (right),~\ref{fig:BRee-et} (left, right), and \ref{fig:BReu-uu}
%(left, right), respectively.
% 
 %Therefore, if the mass of the doubly charged Higgs is located outside of the lowest region, our relations of Eqs.~(\ref{C1})-(\ref{C3}) may remain only academic interest. 
%We may use the more accurate data from other experiments such as oscillations, tritium decays, $0\nu\beta\beta$, and cosmology survey {\it et. al.} in foreseeable future and combine the expressions of Eqs.~(\ref{BRee})-(\ref{Mphase}) to probe the properties of neutrinos.} 
%  

\section{Conclusion}
We have  studied  the close relationships between the neutrino masses and the same sign dilepton decays of the doubly charged Higgs scalars in the model in which 
the  neutrinos are  Majorana particles with their masses are generated radiatively at two-loop level.
Since the dilepton modes in our model could be reachable at the LHC, it is natural to use their branching ratios to infer the neutrino masses.
We have explicitly shown that in the limit of the tribimaximal mixings, the absolute scale of neutrino masses can be expressed in terms of $C_{i}$ (i = 1-3) based on the certain combinations of dilepton branching ratios. It is possible to determine the neutrino masses by just counting the events arising from the dilepton decays of the doubly charged Higgs in its lowest mass region. The allowed parameter space of the fractions among each dilepton branching ratio as the function of the lightest neutrino mass $m_{1}$ is presented. These relations combined with the data from other neutrino experiments may help to set a limit of $m_{1}$ and Majorana phase $\psi_{1,2}$ in the future.  
%  as well as the other absolute neutrino masses and Majorana phases
 %  through the  quantities $C_{i(i = 1 \sim 4)}$ which are composed of different linear combinations of the dilepton branching ratios to be measured at
 %   colliders.\\
    \\
    
    \noindent {\bf Acknowledgments}

This work is supported in part by the National Science Council of ROC under 
Grant \#s: NSC97-2112-M-006-004-MY3 (CSC), NSC-95-2112-M-007-059-MY3 (CQG) and NSC-98-2112-M-007-008-MY3 (CQG)
and by the Boost Program of NTHU (CQG).

\end{document}